\pdfoutput=1
\documentclass[aps,prl,showpacs,twocolumn,letterpaper,superscriptaddress,floatfix]{revtex4-1}
\usepackage{amsmath, amssymb,bm}
\usepackage{natbib}
\usepackage{graphicx}
\usepackage{epsfig}
\usepackage{dcolumn}
\newcolumntype{d}[1]{D{.}{.}{#1}}
\usepackage{array}
\usepackage{tabularx}
\newcolumntype{L}[1]{>{\hsize=#1\hsize\raggedright\arraybackslash}X}
\newcolumntype{R}[1]{>{\hsize=#1\hsize\raggedleft\arraybackslash}X}
\newcolumntype{C}[1]{>{\hsize=#1\hsize\centering\arraybackslash}X}

\newcommand{\be}{\begin{equation}} 
\newcommand{\ee}{\end{equation}}
\newcommand{\ba}{\begin{eqnarray}} 
\newcommand{\ea}{\end{eqnarray}}

\newcommand{\ssec}[1]{\emph{#1}.---}
\begin{document}

\title{Constraining the nuclear energy density functional with quantum Monte Carlo calculations}

\author{Alessandro Roggero}
\email[]{roggero@uw.edu}
\affiliation{Institute for Nuclear Theory, University of Washington, Seattle, WA 98195, US}

\author{Abhishek Mukherjee}
\affiliation{ECT*, Villa Tambosi, I-38123 Villazzano (Trento), Italy}
\affiliation{ClusterVision B.V., Nieuw-Zeelandweg 15B,  1045 AL, Amsterdam, Netherland}

\author{Francesco Pederiva}
\affiliation{Physics Department, University of Trento, via Sommarive 14, I-38123 Trento, Italy}
\affiliation{INFN-TIFPA, Trento Institute for Fundamental Physics and Applications}

\date{\today}

\begin{abstract}
We study the problem of an impurity in fully polarized (spin-up) low density neutron matter with the 
help of an accurate quantum Monte Carlo method in conjunction with a realistic nucleon-nucleon interaction
derived from chiral effective field theory at next-to-next-to-leading-order. Our calculations show that the 
behavior of the proton spin-down impurity is very similar to that of a polaron in a fully polarized unitary 
Fermi gas. We show that our results can be used to put tight constraints on the time-odd parts of the 
energy density functional, independent of the time-even parts, in the density regime relevant to neutron-rich
nuclei and compact astrophysical objects such as neutron stars and supernovae. 
\end{abstract}


\maketitle
\ssec{Introduction} The \emph{ab initio} prediction of nuclear properties from quantum chromodynamics (QCD) remains an 
unresolved challenge in fundamental science. Its importance extends well beyond the confines of basic nuclear physics, 
into the realm of astrophysics, viz. in the physics of neutron stars and core-collapse supernovae.

It is unlikely that direct lattice QCD calculations of many hadron properties will be possible in the forseeable future.
 However, in the past two decades a promising alternative route  has been proposed and pursued with vigor. This scheme consists
 of bridging the gap between QCD and low energy nuclear physics by building successive effective theories. 

In the first step one constructs an effective Hamlitonian with the hadronic degrees
of freedom. The structure of this Hamiltonian is tightly constrained by chiral effective field 
theory (EFT) \cite{Epelbaum2009,*Machleidt2011,*Hammer2013}. 
In the next stage one performs accurate many body calculations with this effective Hamiltonian for simple configurations, 
e.g. homogeneous matter, light and medium mass nuclei etc.
Results from these calculations, in conjunction with experimental data, are eventually used to construct an energy density 
functional (EDF) for nuclear systems. Density
functional theory is, presently, the only viable computational method for complex inhomogeneous systems.

It is of paramount importance that the effective theory at each stage is consistent with the available experimental data and
 the predictions of the underlying microscopic 
theory. A successful prototype is provided by the density functional theory for electronic structure calculations \cite{Perdew1992}
which was fit to accurate quantum Monte Carlo (QMC) calculations for the electron gas \cite{Ceperley1980}. 
Of course, nuclear systems are far more complicated because 
of the complexity of the nuclear forces and the remaining ambiguities  in their short range structure.

Most nuclear EDFs are fit to the ground state properties of even-even nuclei, saturation properties of nuclear matter
and occasionally to microscopic calculations of unpolarized neutron matter. These quantities constrain only that part
of the EDF which depends on the time-reversal-even densities (``time-even part"). The EDF also depends on time-reversal-odd
densities (``time-odd part") which plays an important role in a variety of phenomena:  binding energies of odd-mass nuclei~\cite{Satula1998},
pairing correlations in nuclei~\cite{Duguet2001}, distribution of the Gamow-Teller strength~\cite{Bender2002}, 
properties of rotating nuclei~\cite{Dobaczewski1995,*Post1985}, nuclear magnetism~\cite{Afanasjev2000} etc. At present,
the time-odd part of the Skyrme and other non--relativistic nuclear EDF is ill-determined because of the lack of unambiguous constraints
from experiment or ab--initio calculations.

In the recent past, there is an emerging consensus that the theoretical uncertainities of the nuclear forces is largely suppressed
in low density neutron matter (densities sufficiently less than the saturation density of nuclear matter).  
In this regime, the properties of the relevant components of the two nucleon forces are well established and the 
contributions from three and higher body forces are rather small. 
Any realistic nucleon-nucleon interaction, which fits the low energy nucleon-nucleon scattering phase shifts and the
binding energy of deuteron, in conjunction with an accurate many body method produce consistent ``theoretical data"; which can    
provide constraints for the EDF complementary to those coming from experiments.

In this paper we report the results from fully non-perturbative QMC calculations with a chiral EFT 
Hamiltonian for fully polarized (spin up) low density neutron matter with an impurity (spin down neutron or spin up/down proton). 
The impurity problem that we discuss here is a generalization of the well known polaron problem in solid state systems and in ultracold gases 
(see, e.g. in~\cite{Chevy2010,*Massignan2014}). 
In fact, we find that the proton spin-down impurity behaves in a manner which is qualitatively very similar to a polaron in a fully polarized 
Fermi gas in the unitary regime, i.e., the regime with diverging s-wave scattering length ($a_s \to \infty$) and vanishing effective range ($r_e \to 0$), over
a wide density range $10^{-3} \mbox{ fm}^{-3} \le \rho \le 5\times10^{-2} \mbox{ fm}^{-3}$.

We show that the difference between energies of the proton spin up and spin down impurities depends only on the time-odd 
part of the EDF. Thus, our results provide stringent constraints for   
the time-odd part of the density functional, \emph{independent of the time-even part}.
The results presented here will provide valuable guidance in constructing EDFs in regimes relevant to neutron-rich nuclei, 
neutron star crusts and supernovae neutrinosphere.

\ssec{Method} Our calculations are based on the recently developed QMC method called  the configuration interaction Monte Carlo (CIMC) 
method~\cite{Mukherjee2013,Roggero2013,Roggero2014}.
The CIMC method is based on filtering out an eigenstate $\Psi_0$ of the Hamiltonian $H$  by
repeated application of the propagator $\mathcal{P}=e^{-\tau(H-E_T)}$ on an initial state $\Psi_{\rm I}$,
\be
|\Psi_0 \rangle = \lim_{\mathcal{N}_{\tau} \to \infty} \mathcal{P}^{\mathcal{N}_{\tau}} |\Psi_{\rm I} \rangle . 
\ee
Here, $E_T$ is an energy shift used to keep the norm of the wave
function approximately constant, and $\tau$ is a finite step in `imaginary' time $\tau=it$.
The state, $\Psi_0$, is the eigenstate with the lowest eigenvalue within the subset of states having 
non-zero overlaps with $\Psi_{\rm I}$.

The application of the propagator is carried out stochastically.
The main difference between the CIMC method and traditional continuum diffusion Monte Carlo methods
is that in the CIMC method this stochastic projection is performed in Fock space (
i.e. the basis is provided by the  Slater determinants that can 
constructed from  a finite set of single particle (sp) basis states), as 
opposed to the coordinate space. As a result, non-local Hamiltonians do not pose any technical 
problems.

In this work, we use the sp basis given by  eigenstates of momentum and the $z$ components of spin and isospin.
The calculations for fully polarized neutrons are performed in a box containing $N$ spin-up neutrons. The impurity 
system contains an additional impurity particle.
Periodic boundary conditions are imposed. 
The size of the box is given by the density, $\rho$, of the spin-up neutrons, $L^3=N/\rho$.
The finite size of the box implies that the sp states are restricted to a lattice in momentum space with a lattice constant $l=2\pi/L$. 

A finite sp basis is chosen by imposing a ``basis cutoff" $k_{\rm max}$, so that
only those sp states with $\mathbf{k}^2 \leq k_{\rm max}^2$ are included. 
A sequence of calculations, with successively larger values of $k_{\rm max}$, are  performed till convergence is reached.
We deem the calculations to have converged in $k_{\rm max}$ when the difference in the energies
between the successive calculations are smaller than the statistical error ($\sim 10$ KeV ).

Sampling of new states can be performed under the condition that the matrix elements of the
propagator, $\mathcal{P}$, are always positive semi-definite.
For fermions interacting via realistic potentials, this condition is never fullfilled.
(An interesting exception is provided by the pure pairing Hamiltonian \cite{Cerf1993,*Mukherjee2011}.)
This gives rise to the so-called sign-problem, which we circumvent  
by using a guiding wave function to constrain the random walk to a subsector of the 
full many-body Hilbert space in which the sampling procedure is well defined \cite{Mukherjee2013}. 
This restriction of the random walk introduces an approximation which is similar
to the fixed-node/fixed-phase approximation commonly used in continuum QMC. Our method
provides strict variational upper bounds for the energy.

As explained in Refs.~\onlinecite{Roggero2013, Roggero2014}, we use coupled cluster double (CCD) type wave functions 
as the guiding wave functions. As a result, the CIMC method provides an interesting synthesis of QMC methods and coupled cluster (CC) theory.
In principle, the fixed phase approximation can be systematically improved by including irreducible triples, quadrupoles etc.
in the guiding wave function. However, as discussed in Ref.~\onlinecite{Roggero2014} these contributions are expected to be 
rather small at these densities (less than a few percent of the total correlation energy). 

\begin{figure}
\includegraphics[width=\columnwidth]{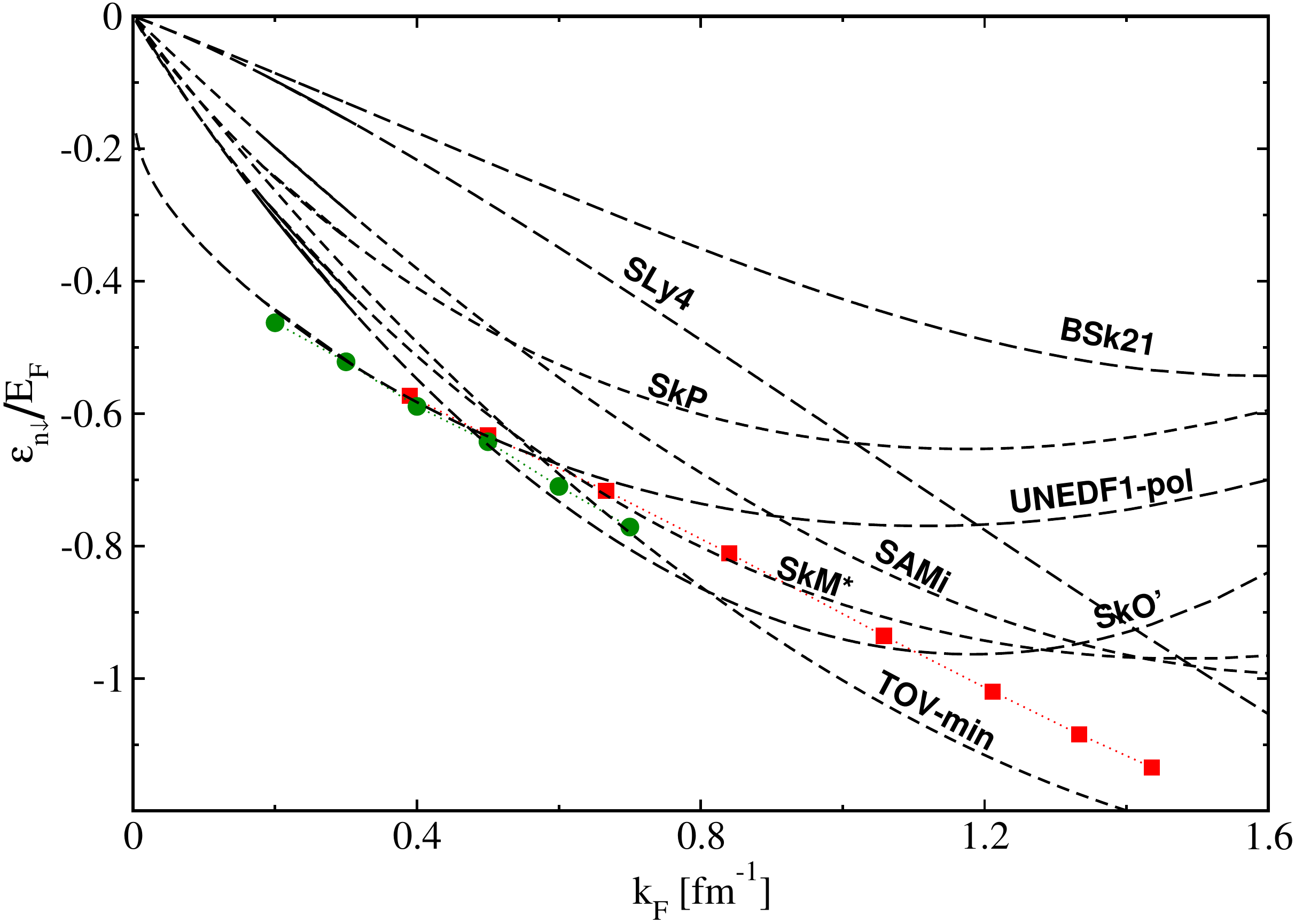}
\caption{\label{n_dn} (Color online) The energy of the neutron spin-down impurity in the units 
of the Fermi energy of the spin-up neutrons. The red filled squares are our QMC results with the NNLO$_{\rm opt}$ 
interaction. The green filled circles are the GFMC calculations with an s-wave interaction fit to the $nn$ scattering 
length and effective range~\cite{Forbes2014}. The black dashed lines are predictions from various density functionals 
(see text).}
\end{figure}

\ssec{Results}
We calculate the ground state energies for a fully polarized system  
and that with an additional impurity particle. The difference between these two energies gives the impurity 
energy.
We use the recently developed next-to-next-to-leading order chiral NNLO$_{\rm opt}$ interaction \cite{Ekstrom2013} 
for our calculations.
The scattering phase shifts obtained from this interaction fit the experimental database \cite{Stoks1993} 
at $\chi^2 \sim 1$ for laboratory energies less than $125$ MeV. However, as alluded to in the introduction,
the conclusions we present \emph{are independent} of the particular interaction model we are using.

In Fig.~\ref{n_dn} we plot the ratio of the energy of neutron spin-down impurity, $\varepsilon_{n\downarrow}$,
 and the Fermi energy of the fully polarized system, $E_F$, 
versus the Fermi momentum $k_{F}$. 
Our results are good agreement with the GFMC calculations reported in Ref.~\onlinecite{Forbes2014} using  
an s-wave interaction (fit to the $nn$ scattering length and effective range).
For example, at $k_F=0.4$ fm$^{-1}$, we get $\varepsilon_{n\uparrow}/E_F=-0.582 \pm 0.002$ while the GFMC calculation gives $-0.589 \pm 0.005$. 
An AFDMC calculation performed the Argonne $v_8'$ potential gives $-0.567 \pm 0.006$ at the same $k_F$.  

\begin{figure}
\includegraphics[width=\columnwidth]{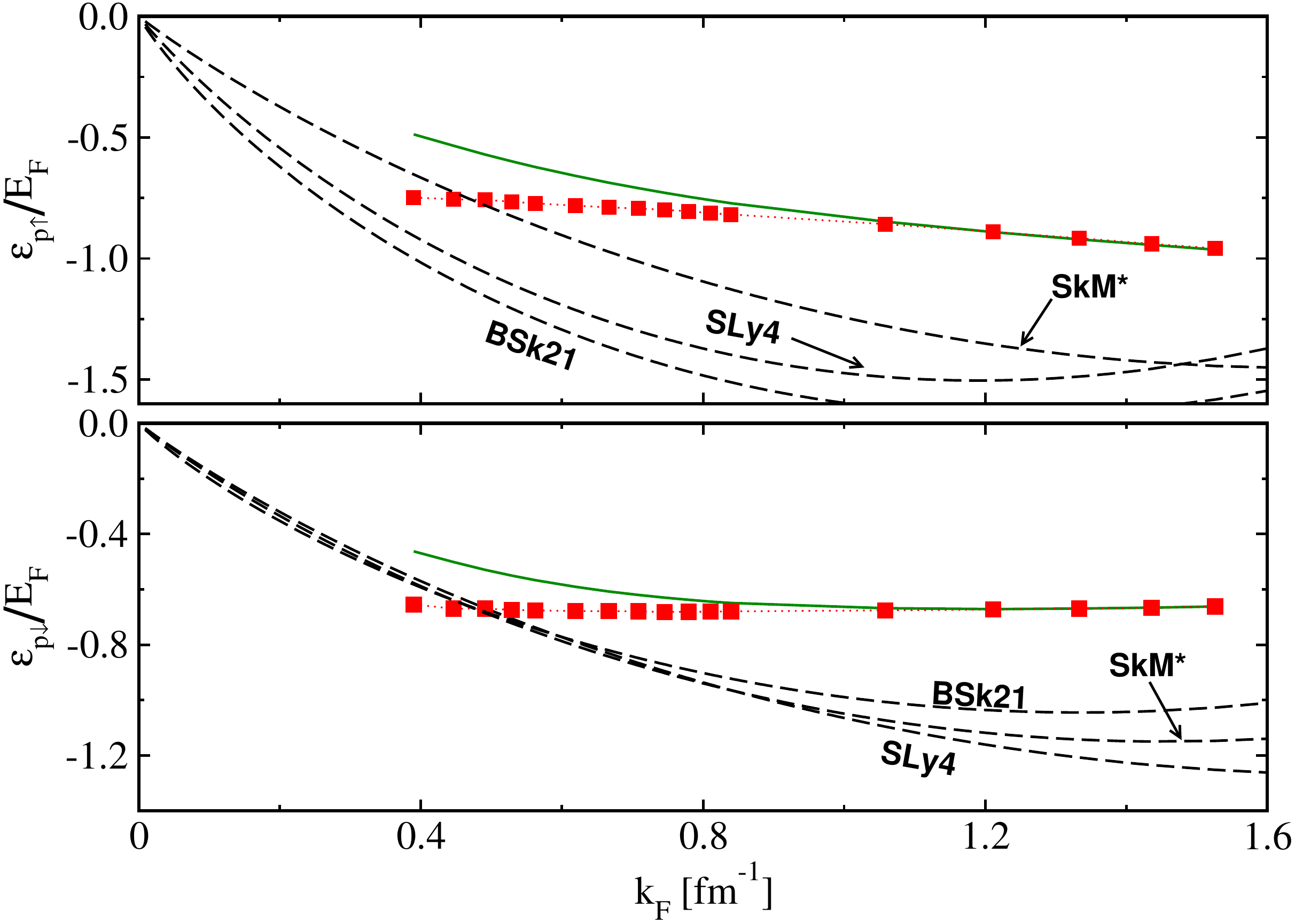}
\caption{\label{p_updn} (Color online) The energy of the proton spin-up (top panel) and spin-down (bottom panel)
 impurities  in the units of the Fermi energy of the spin-up neutrons. The red filled squares are our QMC results with the NNLO$_{\rm opt}$. 
interaction. The green solid lines are the results from second order perturbation theory. 
The black dashed lines are predictions from various density functionals (see text).}
\end{figure}

The impurity energies reported in Fig.~\ref{n_dn} and later in Fig.~\ref{p_updn} 
were performed with $N=7$ spin-up neutrons. We have checked in selected cases that 
the difference between the $N=7$ and the $N=33$ energies is about $1-2$\%.
For example, for $\rho=0.04$ fm$^{-3}$ $\varepsilon_{n\downarrow}/E_F$ is $-0.6698 \pm 0.0005$ 
with $N=7$ and is $-0.664 \pm 0.006$ for $N=33$, while for $\rho = 0.06$ fm$^{-3}$ the corresponding
values are $-0.6617 \pm 0.0003$ and $-0.647 \pm 0.004$. 
With $N=7$ the size of the box, $L$, for the largest density we consider in this work ($\rho=0.06$ fm$^{-3}$)
is about $4.9$ fm. This is about three times the characteristic range of the nucleon-nucleon interaction given by
the pion Compton wavelength ($\approx 1.4$ fm). At higher densities ($\rho \geq 0.16$ fm$^{-3}$) the corrections resulting from
performing calculations with a finite number of particles is expected to sizeable and it is customary to 
perform calculations with larger particle number ($N \geq 33$, for each spin). However, at the densities 
we are considering in this paper, the finite particle number corrections (even at $N=7$) can be reasonably 
expected to be smaller than, or at most comparable to, the other sources of uncertainty (the non inclusion of three body forces
in the Hamiltonian or the absence of triples in the wave function).

In Fig.~(\ref{p_updn}) we plot the ratio of the energy of the proton spin up/down impurity ($\varepsilon_{p\uparrow/\downarrow}$) and $E_F$.
The density dependence of $\varepsilon_{p\uparrow/\downarrow}/E_F$ is rather weak.
In fact, the QMC results for $\varepsilon_{p\downarrow}/E_F$
change by less than $2\% \; (-0.681 < \varepsilon_{p\downarrow}/E_F < -0.666)$ when the density changes by more than 
an order of magnitude ($10^{-3}-5\times10^{-2}$).
Interestingly, this value is larger, in magnitude, than the corresponding (theoretical) value for polaron energy in a fully polarized unitary Fermi gas ($\approx -0.6$)~\cite{Lobo2006,*Prokofev2008}
by about $10\%$. It is worth pointing out here that the singlet $pn$ scattering length is about $25\%$ larger than the singlet $nn$ scattering length.
This weak density dependence of $\varepsilon_{p\uparrow/\downarrow}$ is a non-perturbative result. Calculations from second order perturbation theory, also shown in the 
figure, predict a much stronger density dependence for $k_F < 1.0$ fm.

\begin{figure}
\includegraphics[width=\columnwidth]{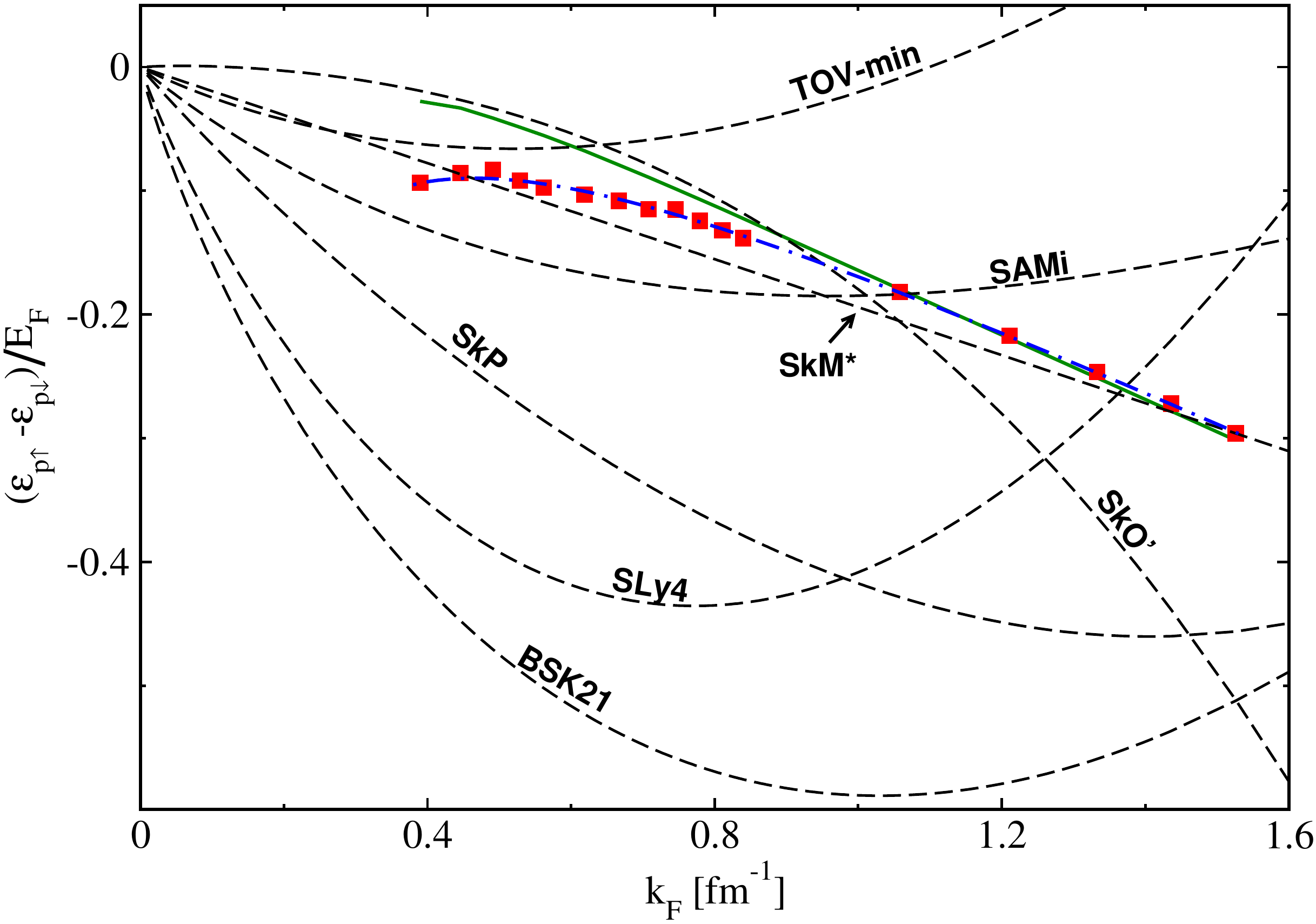}
\caption{\label{diff}(Color online) The difference between the energies of the proton spin-up and spin-down 
 impurities  in the units of the Fermi energy of the spin-up neutrons. The red filled squares are our QMC results with the chiral NNLO$_{\rm opt}$ 
interaction. The green solid line is the prediction from second order perturbation theory. The blue dot-dashed line is a fit of the form : $A-\frac{B}{k_F|a_s|}-C k_Fr_e$.
The black dashed lines are predictions from various density functionals (see text).}
\end{figure}

The Skyrme EDF for uniform matter is usually parametrized as
\be
\label{edf}
\mathcal{E} = \mathcal{E}_{\rm kin} + \sum_{t=0,1} \left ( C^{\rho}_t \rho_{t}^2  + C^{\tau}_{t} \rho_{t} \tau_{t} + C^{s}_{t} s_{t}^2 + C^{T}_{t} s_{t} T_{t} \right ).
\ee
where $\mathcal{E}_{\rm kin}$ is the kinetic energy density. The isoscalar (isovector) density, spin-density, kinetic density 
and spin-kinetic density are denoted by $\rho_0,s_0,\tau_0$ and $T_0$ ($\rho_1,s_1,\tau_1$ and $T_1$ ), respectively. The part 
of the density functional which explicitly depends on the time-odd densities ($s_{t}$, $T_t$) is the time-odd part, and the rest is the time-even part. 

The coefficients $C^{\rho}_{t}, C^{\tau}_{t}, C^s_{t}$ and $C^T_{t}$ can only depend on the total (isoscalar) density $\rho_0=\rho$. In general,
the coefficients are all independent and should be fixed from available data. However, for EDFs derived from a Skyrme force, there are additional
relationships amongst the coefficients and the number of indepedent coefficients is smaller. 
Usually the $C^{\rho}_t$ and $C^{s}_t$ are assumed to have the form
\be
\label{rho-dep}
C^{(\rho/s)}_t = C^{(\rho/s)0}_t + C^{(\rho/s)\rho}_t \rho^{(\gamma/\delta)}.
\ee 

The impurity energy can be calculated from the EDF as
\be
\label{mu}
\varepsilon_{\tau \sigma} = \left .\frac{\partial \mathcal{E}}{\partial \rho_{\tau \sigma}} \right \vert_{\rho_{\tau \sigma} \to 0},
\ee
with $\tau\sigma=\{n\downarrow, p\uparrow, p\downarrow\}$.
In Fig.~(\ref{n_dn}) we also show $\varepsilon_{n\downarrow}/E_F$ obtained
from a wide cross-section of currently popular EDFs: SLy4~\cite{Chabanat1995,*Chabanat1997,*Chabanat1998}, SkM*~\cite{Bartel1982}, BSk21~\cite{Goriely2010},
SkP~\cite{Dobaczewski1984}, SkO$^{\prime}$~\cite{Reinhard1999}, SAMi~\cite{Roca2012}, TOV-min~\cite{Erler2013} and UNEDF-pol~\cite{Forbes2014}. 
In Fig.~\ref{p_updn}, we show $\varepsilon_{p\uparrow/\downarrow}/E_F$ for a smaller sub-section of the EDFs. This is done in order to avoid over-crowding the
figure. However, we would like to note here that the three EDFs, which are plotted in Fig.~\ref{p_updn}, provide a fair representation of the spread
in the predictions from the current Skyrme--type EDFs; all the other EDFs show very similar trends both qualitatively and quantitatively.

None of the EDFs reproduce the QMC results satisfactorily. This is even more evident in the case of the proton spin-down impurity; 
whereas all the EDFs predict $\varepsilon_{p\downarrow}/E_F$ to be decreasing with $k_F$, our QMC calculations predict a flat behavior.
This is not unexpected since the EDFs are usually fit to the experimental properties nuclear systems near saturation density
and low isospin polarization (stable nuclei), and many body calculations of unpolarized neutron matter.
On the basis of our calculations we conclude that in order to account for the correlations in the low density matter in the presence of 
large spin and isospin polarization, \emph{qualitative changes} are warranted in the form of the EDFs.

The difference $\varepsilon_{p\uparrow} - \varepsilon_{p \downarrow}$ is a purely time-odd quantity. 
From Eqs.~(\ref{edf}) and (\ref{mu}) one can easily obtain the following relation 
\be
\frac{\varepsilon_{p\uparrow} - \varepsilon_{p \downarrow}}{E_F} = \frac{4m(C^s_0 - C^s_1)}{3\pi^2 \hbar^2} k_F -  \frac{2m(C^T_0 - C^T_1)}{5\pi^2 \hbar^2} k_F^3.   
\ee
In Fig.(\ref{diff}) we
compare the predictions from our QMC calculations for $(\varepsilon_{p\uparrow} -\varepsilon{p\downarrow})/E_F$ 
with those from different EDFs. It is clear that none of the EDFs correctly describe our results.
The SkM* EDF reproduces the linear part of our results reasonablly well. However, the SkM* EDF does not perform 
any better than the other EDFs for the individual $\varepsilon_{p\uparrow/\downarrow}$. Also, globally the SkM* EDF fares
significantly worse than the more modern EDFs in describing experimental data for nuclei (e.g., masses). 

Our results are well fit by the form
\be
\frac{\varepsilon_{p\uparrow} - \varepsilon_{p \downarrow}}{E_F}=A-\frac{B}{k_F|a_s|}-C k_Fr_e
\ee
with $A=0.17\pm0.01$, $B=1.4\pm0.1$ and $C=0.101\pm0.001$. We have used the values $a_s=-23.75$ fm and $r_e=2.75$ fm for the 
neutron-proton singlet scattering length and effective length, respectively. This form is clearly reminscent of a
dilute unitary Fermi gas.

\ssec{Conclusion} We have presented QMC calculations with a chiral interaction for the impurities in low density fully polarized 
neutron matter. The proton spin-down impurity shows universal behaviour for a wide range of densities. None of the state of the art
Skyrme EDFs describe our microscopic calculations correctly. We showed that the difference between the proton impurity energies
depends only on the time-odd part of the EDF. We found a simple functional form which fits our results for this difference, but is 
nevertheless qualitatively different from what is predicted by the current functional forms used in the Skyrme EDFs. Our results provide 
new constraints for constructing accurate density functionals.  

\ssec{Acknowledgments.} We thank A.~Gezerlis for sharing the results of their numerical simulations.
The authors are also members of LISC, the Interdisciplinary Laboratory 
for Computational Science, a joint venture between Fondazione Bruno Kessler and the University of Trento. 
Computations have been carried out mostly on the open facilities at
Lawrence Livermore National Laboratory.
\bibliography{edf_chiral}

\begin{thebibliography}{32}%
\makeatletter
\providecommand \@ifxundefined [1]{%
 \@ifx{#1\undefined}
}%
\providecommand \@ifnum [1]{%
 \ifnum #1\expandafter \@firstoftwo
 \else \expandafter \@secondoftwo
 \fi
}%
\providecommand \@ifx [1]{%
 \ifx #1\expandafter \@firstoftwo
 \else \expandafter \@secondoftwo
 \fi
}%
\providecommand \natexlab [1]{#1}%
\providecommand \enquote  [1]{``#1''}%
\providecommand \bibnamefont  [1]{#1}%
\providecommand \bibfnamefont [1]{#1}%
\providecommand \citenamefont [1]{#1}%
\providecommand \href@noop [0]{\@secondoftwo}%
\providecommand \href [0]{\begingroup \@sanitize@url \@href}%
\providecommand \@href[1]{\@@startlink{#1}\@@href}%
\providecommand \@@href[1]{\endgroup#1\@@endlink}%
\providecommand \@sanitize@url [0]{\catcode `\\12\catcode `\$12\catcode
  `\&12\catcode `\#12\catcode `\^12\catcode `\_12\catcode `\%12\relax}%
\providecommand \@@startlink[1]{}%
\providecommand \@@endlink[0]{}%
\providecommand \url  [0]{\begingroup\@sanitize@url \@url }%
\providecommand \@url [1]{\endgroup\@href {#1}{\urlprefix }}%
\providecommand \urlprefix  [0]{URL }%
\providecommand \Eprint [0]{\href }%
\providecommand \doibase [0]{http://dx.doi.org/}%
\providecommand \selectlanguage [0]{\@gobble}%
\providecommand \bibinfo  [0]{\@secondoftwo}%
\providecommand \bibfield  [0]{\@secondoftwo}%
\providecommand \translation [1]{[#1]}%
\providecommand \BibitemOpen [0]{}%
\providecommand \bibitemStop [0]{}%
\providecommand \bibitemNoStop [0]{.\EOS\space}%
\providecommand \EOS [0]{\spacefactor3000\relax}%
\providecommand \BibitemShut  [1]{\csname bibitem#1\endcsname}%
\let\auto@bib@innerbib\@empty
\bibitem [{\citenamefont {{Epelbaum}}\ \emph {et~al.}(2009)\citenamefont
  {{Epelbaum}}, \citenamefont {{Hammer}},\ and\ \citenamefont
  {{Mei{\ss}ner}}}]{Epelbaum2009}%
  \BibitemOpen
  \bibfield  {author} {\bibinfo {author} {\bibfnamefont {E.}~\bibnamefont
  {{Epelbaum}}}, \bibinfo {author} {\bibfnamefont {H.~W.}\ \bibnamefont
  {{Hammer}}}, \ and\ \bibinfo {author} {\bibfnamefont {U.~G.}\ \bibnamefont
  {{Mei{\ss}ner}}},\ }\href {\doibase 10.1103/RevModPhys.81.1773} {\bibfield
  {journal} {\bibinfo  {journal} {Rev. Mod. Phys.}\ }\textbf {\bibinfo {volume}
  {81}},\ \bibinfo {pages} {1773} (\bibinfo {year} {2009})}\BibitemShut
  {NoStop}%
\bibitem [{\citenamefont {{Machleidt}}\ and\ \citenamefont
  {{Entem}}(2011)}]{Machleidt2011}%
  \BibitemOpen
  \bibfield  {author} {\bibinfo {author} {\bibfnamefont {R.}~\bibnamefont
  {{Machleidt}}}\ and\ \bibinfo {author} {\bibfnamefont {D.~R.}\ \bibnamefont
  {{Entem}}},\ }\href {\doibase 10.1016/j.physrep.2011.02.001} {\bibfield
  {journal} {\bibinfo  {journal} {Phys. Rep.}\ }\textbf {\bibinfo {volume}
  {503}},\ \bibinfo {pages} {1} (\bibinfo {year} {2011})}\BibitemShut {NoStop}%
\bibitem [{\citenamefont {{Hammer}}\ \emph {et~al.}(2013)\citenamefont
  {{Hammer}}, \citenamefont {{Nogga}},\ and\ \citenamefont
  {{Schwenk}}}]{Hammer2013}%
  \BibitemOpen
  \bibfield  {author} {\bibinfo {author} {\bibfnamefont {H.~W.}\ \bibnamefont
  {{Hammer}}}, \bibinfo {author} {\bibfnamefont {A.}~\bibnamefont {{Nogga}}}, \
  and\ \bibinfo {author} {\bibfnamefont {A.}~\bibnamefont {{Schwenk}}},\ }\href
  {\doibase 10.1103/RevModPhys.85.197} {\bibfield  {journal} {\bibinfo
  {journal} {Rev. Mod. Phys.}\ }\textbf {\bibinfo {volume} {85}},\ \bibinfo
  {pages} {197} (\bibinfo {year} {2013})}\BibitemShut {NoStop}%
\bibitem [{\citenamefont {Perdew}\ and\ \citenamefont
  {Wang}(1992)}]{Perdew1992}%
  \BibitemOpen
  \bibfield  {author} {\bibinfo {author} {\bibfnamefont {J.~P.}\ \bibnamefont
  {Perdew}}\ and\ \bibinfo {author} {\bibfnamefont {Y.}~\bibnamefont {Wang}},\
  }\href {\doibase 10.1103/PhysRevB.45.13244} {\bibfield  {journal} {\bibinfo
  {journal} {Phys. Rev. B}\ }\textbf {\bibinfo {volume} {45}},\ \bibinfo
  {pages} {13244} (\bibinfo {year} {1992})}\BibitemShut {NoStop}%
\bibitem [{\citenamefont {Ceperley}\ and\ \citenamefont
  {Alder}(1980)}]{Ceperley1980}%
  \BibitemOpen
  \bibfield  {author} {\bibinfo {author} {\bibfnamefont {D.M.}~\bibnamefont
  {Ceperley}}\ and\ \bibinfo {author} {\bibfnamefont {B.J.}~\bibnamefont
  {Alder}},\ }\href@noop {} {\bibfield  {journal} {\bibinfo  {journal} {Phys.
  Rev. Lett.}\ }\textbf {\bibinfo {volume} {45}},\ \bibinfo {pages} {566}
  (\bibinfo {year} {1980})}\BibitemShut {NoStop}%
\bibitem [{\citenamefont {Satula}()}]{Satula1998}%
  \BibitemOpen
  \bibfield  {author} {\bibinfo {author} {\bibfnamefont {W.}~\bibnamefont
  {Satula}},\ }in\ \href@noop {} {\emph {\bibinfo {booktitle} {Nuclear
  Structure'98, AIP Conf. Proc. No. 481 (AIP, New York, 1999)}}},\ \bibinfo
  {editor} {edited by\ \bibinfo {editor} {\bibfnamefont {C.}~\bibnamefont
  {Baktash}}},\ p.\ \bibinfo {pages} {114}\BibitemShut {NoStop}%
\bibitem [{\citenamefont {Duguet}\ \emph {et~al.}(2001)\citenamefont {Duguet},
  \citenamefont {Bonche}, \citenamefont {Heenen},\ and\ \citenamefont
  {Meyer}}]{Duguet2001}%
  \BibitemOpen
  \bibfield  {author} {\bibinfo {author} {\bibfnamefont {T.}~\bibnamefont
  {Duguet}}, \bibinfo {author} {\bibfnamefont {P.}~\bibnamefont {Bonche}},
  \bibinfo {author} {\bibfnamefont {P.~H.}\ \bibnamefont {Heenen}}, \ and\
  \bibinfo {author} {\bibfnamefont {J.}~\bibnamefont {Meyer}},\ }\href
  {\doibase 10.1103/PhysRevC.65.014310} {\bibfield  {journal} {\bibinfo
  {journal} {Phys. Rev. C}\ }\textbf {\bibinfo {volume} {65}},\ \bibinfo
  {pages} {014310} (\bibinfo {year} {2001})}\BibitemShut {NoStop}%
\bibitem [{\citenamefont {Bender}\ \emph {et~al.}(2002)\citenamefont {Bender},
  \citenamefont {Dobaczewski}, \citenamefont {Engel},\ and\ \citenamefont
  {Nazarewicz}}]{Bender2002}%
  \BibitemOpen
  \bibfield  {author} {\bibinfo {author} {\bibfnamefont {M.}~\bibnamefont
  {Bender}}, \bibinfo {author} {\bibfnamefont {J.}~\bibnamefont {Dobaczewski}},
  \bibinfo {author} {\bibfnamefont {J.}~\bibnamefont {Engel}}, \ and\ \bibinfo
  {author} {\bibfnamefont {W.}~\bibnamefont {Nazarewicz}},\ }\href {\doibase
  10.1103/PhysRevC.65.054322} {\bibfield  {journal} {\bibinfo  {journal} {Phys.
  Rev. C}\ }\textbf {\bibinfo {volume} {65}},\ \bibinfo {pages} {054322}
  (\bibinfo {year} {2002})}\BibitemShut {NoStop}%
\bibitem [{\citenamefont {Dobaczewski}\ and\ \citenamefont
  {Dudek}(1995)}]{Dobaczewski1995}%
  \BibitemOpen
  \bibfield  {author} {\bibinfo {author} {\bibfnamefont {J.}~\bibnamefont
  {Dobaczewski}}\ and\ \bibinfo {author} {\bibfnamefont {J.}~\bibnamefont
  {Dudek}},\ }\href {\doibase 10.1103/PhysRevC.52.1827} {\bibfield  {journal}
  {\bibinfo  {journal} {Phys. Rev. C}\ }\textbf {\bibinfo {volume} {52}},\
  \bibinfo {pages} {1827} (\bibinfo {year} {1995})}\BibitemShut {NoStop}%
\bibitem [{\citenamefont {Post}\ \emph {et~al.}(1985)\citenamefont {Post},
  \citenamefont {W\"{u}st},\ and\ \citenamefont {Mosel}}]{Post1985}%
  \BibitemOpen
  \bibfield  {author} {\bibinfo {author} {\bibfnamefont {U.}~\bibnamefont
  {Post}}, \bibinfo {author} {\bibfnamefont {E.}~\bibnamefont {W\"{u}st}}, \
  and\ \bibinfo {author} {\bibfnamefont {U.}~\bibnamefont {Mosel}},\ }\href
  {\doibase http://dx.doi.org/10.1016/S0375-9474(85)90089-2} {\bibfield
  {journal} {\bibinfo  {journal} {Nucl. Phys.}\ }\textbf {\bibinfo {volume}
  {A437}},\ \bibinfo {pages} {274} (\bibinfo {year} {1985})}\BibitemShut
  {NoStop}%
\bibitem [{\citenamefont {Afanasjev}\ and\ \citenamefont
  {Ring}(2000)}]{Afanasjev2000}%
  \BibitemOpen
  \bibfield  {author} {\bibinfo {author} {\bibfnamefont {A.~V.}\ \bibnamefont
  {Afanasjev}}\ and\ \bibinfo {author} {\bibfnamefont {P.}~\bibnamefont
  {Ring}},\ }\href {\doibase 10.1103/PhysRevC.62.031302} {\bibfield  {journal}
  {\bibinfo  {journal} {Phys. Rev. C}\ }\textbf {\bibinfo {volume} {62}},\
  \bibinfo {pages} {031302} (\bibinfo {year} {2000})}\BibitemShut {NoStop}%
\bibitem [{\citenamefont {Chevy}\ and\ \citenamefont {Mora}(2010)}]{Chevy2010}%
  \BibitemOpen
  \bibfield  {author} {\bibinfo {author} {\bibfnamefont {F.}~\bibnamefont
  {Chevy}}\ and\ \bibinfo {author} {\bibfnamefont {C.}~\bibnamefont {Mora}},\
  }\href {http://stacks.iop.org/0034-4885/73/i=11/a=112401} {\bibfield
  {journal} {\bibinfo  {journal} {Rept. Prog. Phys.}\ }\textbf {\bibinfo
  {volume} {73}},\ \bibinfo {pages} {112401} (\bibinfo {year}
  {2010})}\BibitemShut {NoStop}%
\bibitem [{\citenamefont {Massignan}\ \emph {et~al.}(2014)\citenamefont
  {Massignan}, \citenamefont {Zaccanti},\ and\ \citenamefont
  {Bruun}}]{Massignan2014}%
  \BibitemOpen
  \bibfield  {author} {\bibinfo {author} {\bibfnamefont {P.}~\bibnamefont
  {Massignan}}, \bibinfo {author} {\bibfnamefont {M.}~\bibnamefont {Zaccanti}},
  \ and\ \bibinfo {author} {\bibfnamefont {G.~M.}\ \bibnamefont {Bruun}},\
  }\href {http://stacks.iop.org/0034-4885/77/i=3/a=034401} {\bibfield
  {journal} {\bibinfo  {journal} {Reports on Progress in Physics}\ }\textbf
  {\bibinfo {volume} {77}},\ \bibinfo {pages} {034401} (\bibinfo {year}
  {2014})}\BibitemShut {NoStop}%
\bibitem [{\citenamefont {Mukherjee}\ and\ \citenamefont
  {Alhassid}(2013)}]{Mukherjee2013}%
  \BibitemOpen
  \bibfield  {author} {\bibinfo {author} {\bibfnamefont {A.}~\bibnamefont
  {Mukherjee}}\ and\ \bibinfo {author} {\bibfnamefont {Y.}~\bibnamefont
  {Alhassid}},\ }\href {\doibase 10.1103/PhysRevA.88.053622} {\bibfield
  {journal} {\bibinfo  {journal} {Phys. Rev. A}\ }\textbf {\bibinfo {volume}
  {88}},\ \bibinfo {pages} {053622} (\bibinfo {year} {2013})}\BibitemShut
  {NoStop}%
\bibitem [{\citenamefont {Roggero}\ \emph {et~al.}(2013)\citenamefont
  {Roggero}, \citenamefont {Mukherjee},\ and\ \citenamefont
  {Pederiva}}]{Roggero2013}%
  \BibitemOpen
  \bibfield  {author} {\bibinfo {author} {\bibfnamefont {A.}~\bibnamefont
  {Roggero}}, \bibinfo {author} {\bibfnamefont {A.}~\bibnamefont {Mukherjee}},
  \ and\ \bibinfo {author} {\bibfnamefont {F.}~\bibnamefont {Pederiva}},\
  }\href {\doibase 10.1103/PhysRevB.88.115138} {\bibfield  {journal} {\bibinfo
  {journal} {Phys. Rev. B}\ }\textbf {\bibinfo {volume} {88}},\ \bibinfo
  {pages} {115138} (\bibinfo {year} {2013})}\BibitemShut {NoStop}%
\bibitem [{\citenamefont {Roggero}\ \emph {et~al.}(2014)\citenamefont
  {Roggero}, \citenamefont {Mukherjee},\ and\ \citenamefont
  {Pederiva}}]{Roggero2014}%
  \BibitemOpen
  \bibfield  {author} {\bibinfo {author} {\bibfnamefont {A.}~\bibnamefont
  {Roggero}}, \bibinfo {author} {\bibfnamefont {A.}~\bibnamefont {Mukherjee}},
  \ and\ \bibinfo {author} {\bibfnamefont {F.}~\bibnamefont {Pederiva}},\
  }\href@noop {} {\bibfield  {journal} {\bibinfo  {journal} {Phys. Rev. Lett.}\
  \textbf {\bibinfo {volume}{112}},\ \bibinfo {pages} {221103}} (\bibinfo {year} {2014})},\ \Eprint
  {http://arxiv.org/abs/arXiv:1402.1576~[nucl-th]} {arXiv:1402.1576~[nucl-th]}
  \BibitemShut {NoStop}%
\bibitem [{\citenamefont {Cerf}\ and\ \citenamefont {Martin}(1993)}]{Cerf1993}%
  \BibitemOpen
  \bibfield  {author} {\bibinfo {author} {\bibfnamefont {N.}~\bibnamefont
  {Cerf}}\ and\ \bibinfo {author} {\bibfnamefont {O.}~\bibnamefont {Martin}},\
  }\href@noop {} {\bibfield  {journal} {\bibinfo  {journal} {Phys. Rev. C}\
  }\textbf {\bibinfo {volume} {47}},\ \bibinfo {pages} {2610} (\bibinfo {year}
  {1993})}\BibitemShut {NoStop}%
\bibitem [{\citenamefont {Mukherjee}\ \emph {et~al.}(2011)\citenamefont
  {Mukherjee}, \citenamefont {Alhassid},\ and\ \citenamefont
  {Bertsch}}]{Mukherjee2011}%
  \BibitemOpen
  \bibfield  {author} {\bibinfo {author} {\bibfnamefont {A.}~\bibnamefont
  {Mukherjee}}, \bibinfo {author} {\bibfnamefont {Y.}~\bibnamefont {Alhassid}},
  \ and\ \bibinfo {author} {\bibfnamefont {G.F.}~\bibnamefont {Bertsch}},\
  }\href@noop {} {\bibfield  {journal} {\bibinfo  {journal} {Phys. Rev. C}\
  }\textbf {\bibinfo {volume} {83}},\ \bibinfo {pages} {014319} (\bibinfo
  {year} {2011})}\BibitemShut {NoStop}%
\bibitem [{\citenamefont {Forbes}\ \emph {et~al.}(2014)\citenamefont {Forbes},
  \citenamefont {Gezerlis}, \citenamefont {Hebeler}, \citenamefont {Lesinski},\
  and\ \citenamefont {Schwenk}}]{Forbes2014}%
  \BibitemOpen
  \bibfield  {author} {\bibinfo {author} {\bibfnamefont {M.M.}~\bibnamefont
  {Forbes}}, \bibinfo {author} {\bibfnamefont {A.}~\bibnamefont {Gezerlis}},
  \bibinfo {author} {\bibfnamefont {K.}~\bibnamefont {Hebeler}}, \bibinfo
  {author} {\bibfnamefont {T.}~\bibnamefont {Lesinski}}, \ and\ \bibinfo
  {author} {\bibfnamefont {A.}~\bibnamefont {Schwenk}},\ }\href@noop {}
  {\bibfield  {journal} {\bibinfo  {journal} {Phys. Rev. C}\ }\textbf {\bibinfo
  {volume} {89}},\ \bibinfo {pages} {041301(R)} (\bibinfo {year}
  {2014})}\BibitemShut {NoStop}%
\bibitem [{\citenamefont {Ekstr\"{o}m}\ \emph {et~al.}(2013)\citenamefont
  {Ekstr\"{o}m}, \citenamefont {Baardsen}, \citenamefont {Forss\'{e}n},
  \citenamefont {Hagen}, \citenamefont {Hjorth-Jensen}, \citenamefont {Jansen},
  \citenamefont {Machleidt}, \citenamefont {Nazarewicz}, \citenamefont
  {Papenbrock}, \citenamefont {Sarich},\ and\ \citenamefont
  {Wild}}]{Ekstrom2013}%
  \BibitemOpen
  \bibfield  {author} {\bibinfo {author} {\bibfnamefont {A.}~\bibnamefont
  {Ekstr\"{o}m}}, \bibinfo {author} {\bibfnamefont {G.}~\bibnamefont
  {Baardsen}}, \bibinfo {author} {\bibfnamefont {C.}~\bibnamefont
  {Forss\'{e}n}}, \bibinfo {author} {\bibfnamefont {G.}~\bibnamefont {Hagen}},
  \bibinfo {author} {\bibfnamefont {M.}~\bibnamefont {Hjorth-Jensen}}, \bibinfo
  {author} {\bibfnamefont {G.~R.}\ \bibnamefont {Jansen}}, \bibinfo {author}
  {\bibfnamefont {R.}~\bibnamefont {Machleidt}}, \bibinfo {author}
  {\bibfnamefont {W.}~\bibnamefont {Nazarewicz}}, \bibinfo {author}
  {\bibfnamefont {T.}~\bibnamefont {Papenbrock}}, \bibinfo {author}
  {\bibfnamefont {J.}~\bibnamefont {Sarich}}, \ and\ \bibinfo {author}
  {\bibfnamefont {S.~M.}\ \bibnamefont {Wild}},\ }\href {\doibase
  10.1103/PhysRevLett.110.192502} {\bibfield  {journal} {\bibinfo  {journal}
  {Phys. Rev. Lett.}\ }\textbf {\bibinfo {volume} {110}},\ \bibinfo {pages}
  {192502} (\bibinfo {year} {2013})}\BibitemShut {NoStop}%
\bibitem [{\citenamefont {Stoks}\ \emph {et~al.}(1993)\citenamefont {Stoks},
  \citenamefont {Klomp}, \citenamefont {Rentmeester},\ and\ \citenamefont
  {de~Swart}}]{Stoks1993}%
  \BibitemOpen
  \bibfield  {author} {\bibinfo {author} {\bibfnamefont {V.~G.~J.}\
  \bibnamefont {Stoks}}, \bibinfo {author} {\bibfnamefont {R.~A.~M.}\
  \bibnamefont {Klomp}}, \bibinfo {author} {\bibfnamefont {M.~C.~M.}\
  \bibnamefont {Rentmeester}}, \ and\ \bibinfo {author} {\bibfnamefont {J.~J.}\
  \bibnamefont {de~Swart}},\ }\href {\doibase 10.1103/PhysRevC.48.792}
  {\bibfield  {journal} {\bibinfo  {journal} {Phys. Rev. C}\ }\textbf {\bibinfo
  {volume} {48}},\ \bibinfo {pages} {792} (\bibinfo {year} {1993})}\BibitemShut
  {NoStop}%
\bibitem [{\citenamefont {Lobo}\ \emph {et~al.}(2006)\citenamefont {Lobo},
  \citenamefont {Recati}, \citenamefont {Giorgini},\ and\ \citenamefont
  {Stringari}}]{Lobo2006}%
  \BibitemOpen
  \bibfield  {author} {\bibinfo {author} {\bibfnamefont {C.}~\bibnamefont
  {Lobo}}, \bibinfo {author} {\bibfnamefont {A.}~\bibnamefont {Recati}},
  \bibinfo {author} {\bibfnamefont {S.}~\bibnamefont {Giorgini}}, \ and\
  \bibinfo {author} {\bibfnamefont {S.}~\bibnamefont {Stringari}},\ }\href
  {\doibase 10.1103/PhysRevLett.97.200403} {\bibfield  {journal} {\bibinfo
  {journal} {Phys. Rev. Lett.}\ }\textbf {\bibinfo {volume} {97}},\ \bibinfo
  {pages} {200403} (\bibinfo {year} {2006})}\BibitemShut {NoStop}%
\bibitem [{\citenamefont {Prokof'ev}\ and\ \citenamefont
  {Svistunov}(2008)}]{Prokofev2008}%
  \BibitemOpen
  \bibfield  {author} {\bibinfo {author} {\bibfnamefont {N.}~\bibnamefont
  {Prokof'ev}}\ and\ \bibinfo {author} {\bibfnamefont {B.}~\bibnamefont
  {Svistunov}},\ }\href {\doibase 10.1103/PhysRevB.77.020408} {\bibfield
  {journal} {\bibinfo  {journal} {Phys. Rev. B}\ }\textbf {\bibinfo {volume}
  {77}},\ \bibinfo {pages} {020408} (\bibinfo {year} {2008})}\BibitemShut
  {NoStop}%
\bibitem [{\citenamefont {Chabanat}\ \emph {et~al.}(1995)\citenamefont
  {Chabanat}, \citenamefont {Bonche}, \citenamefont {Haensel}, \citenamefont
  {Meyer},\ and\ \citenamefont {Schaeffer}}]{Chabanat1995}%
  \BibitemOpen
  \bibfield  {author} {\bibinfo {author} {\bibfnamefont {E.}~\bibnamefont
  {Chabanat}}, \bibinfo {author} {\bibfnamefont {P.}~\bibnamefont {Bonche}},
  \bibinfo {author} {\bibfnamefont {P.}~\bibnamefont {Haensel}}, \bibinfo
  {author} {\bibfnamefont {J.}~\bibnamefont {Meyer}}, \ and\ \bibinfo {author}
  {\bibfnamefont {R.}~\bibnamefont {Schaeffer}},\ }\href@noop {} {\bibfield
  {journal} {\bibinfo  {journal} {Physica Scripta}\ }\textbf {\bibinfo {volume}
  {1995}},\ \bibinfo {pages} {231} (\bibinfo {year} {1995})}\BibitemShut
  {NoStop}%
\bibitem [{\citenamefont {Chabanat}\ \emph {et~al.}(1997)\citenamefont
  {Chabanat}, \citenamefont {Bonche}, \citenamefont {Haensel}, \citenamefont
  {Meyer},\ and\ \citenamefont {Schaeffer}}]{Chabanat1997}%
  \BibitemOpen
  \bibfield  {author} {\bibinfo {author} {\bibfnamefont {E.}~\bibnamefont
  {Chabanat}}, \bibinfo {author} {\bibfnamefont {P.}~\bibnamefont {Bonche}},
  \bibinfo {author} {\bibfnamefont {P.}~\bibnamefont {Haensel}}, \bibinfo
  {author} {\bibfnamefont {J.}~\bibnamefont {Meyer}}, \ and\ \bibinfo {author}
  {\bibfnamefont {R.}~\bibnamefont {Schaeffer}},\ }\href@noop {} {\bibfield
  {journal} {\bibinfo  {journal} {Nucl. Phys.}\ }\textbf {\bibinfo {volume}
  {A627}},\ \bibinfo {pages} {710} (\bibinfo {year} {1997})}\BibitemShut
  {NoStop}%
\bibitem [{\citenamefont {Chabanat}\ \emph {et~al.}(1998)\citenamefont
  {Chabanat}, \citenamefont {Bonche}, \citenamefont {Haensel}, \citenamefont
  {Meyer},\ and\ \citenamefont {Schaeffer}}]{Chabanat1998}%
  \BibitemOpen
  \bibfield  {author} {\bibinfo {author} {\bibfnamefont {E.}~\bibnamefont
  {Chabanat}}, \bibinfo {author} {\bibfnamefont {P.}~\bibnamefont {Bonche}},
  \bibinfo {author} {\bibfnamefont {P.}~\bibnamefont {Haensel}}, \bibinfo
  {author} {\bibfnamefont {J.}~\bibnamefont {Meyer}}, \ and\ \bibinfo {author}
  {\bibfnamefont {R.}~\bibnamefont {Schaeffer}},\ }\href@noop {} {\bibfield
  {journal} {\bibinfo  {journal} {Nucl. Phys.}\ }\textbf {\bibinfo {volume}
  {A635}},\ \bibinfo {pages} {231} (\bibinfo {year} {1998})}\BibitemShut
  {NoStop}%
\bibitem [{\citenamefont {Bartel}\ \emph {et~al.}(1982)\citenamefont {Bartel},
  \citenamefont {Quentin}, \citenamefont {Brack}, \citenamefont {Guet},\ and\
  \citenamefont {H{\aa}kansson}}]{Bartel1982}%
  \BibitemOpen
  \bibfield  {author} {\bibinfo {author} {\bibfnamefont {J.}~\bibnamefont
  {Bartel}}, \bibinfo {author} {\bibfnamefont {P.}~\bibnamefont {Quentin}},
  \bibinfo {author} {\bibfnamefont {M.}~\bibnamefont {Brack}}, \bibinfo
  {author} {\bibfnamefont {C.}~\bibnamefont {Guet}}, \ and\ \bibinfo {author}
  {\bibfnamefont {H.-B.}\ \bibnamefont {H{\aa}kansson}},\ }\href@noop {}
  {\bibfield  {journal} {\bibinfo  {journal} {Nucl. Phys.}\ }\textbf {\bibinfo
  {volume} {A386}},\ \bibinfo {pages} {79} (\bibinfo {year}
  {1982})}\BibitemShut {NoStop}%
\bibitem [{\citenamefont {Goriely}\ \emph {et~al.}(2010)\citenamefont
  {Goriely}, \citenamefont {Chamel},\ and\ \citenamefont
  {Pearson}}]{Goriely2010}%
  \BibitemOpen
  \bibfield  {author} {\bibinfo {author} {\bibfnamefont {S.}~\bibnamefont
  {Goriely}}, \bibinfo {author} {\bibfnamefont {N.}~\bibnamefont {Chamel}}, \
  and\ \bibinfo {author} {\bibfnamefont {J.M.}~\bibnamefont {Pearson}},\
  }\href@noop {} {\bibfield  {journal} {\bibinfo  {journal} {Phys. Rev. C}\
  }\textbf {\bibinfo {volume} {82}},\ \bibinfo {pages} {035804} (\bibinfo
  {year} {2010})}\BibitemShut {NoStop}%
\bibitem [{\citenamefont {Dobaczewski}\ \emph {et~al.}(1984)\citenamefont
  {Dobaczewski}, \citenamefont {Flocard},\ and\ \citenamefont
  {Treiner}}]{Dobaczewski1984}%
  \BibitemOpen
  \bibfield  {author} {\bibinfo {author} {\bibfnamefont {J.}~\bibnamefont
  {Dobaczewski}}, \bibinfo {author} {\bibfnamefont {H.}~\bibnamefont
  {Flocard}}, \ and\ \bibinfo {author} {\bibfnamefont {J.}~\bibnamefont
  {Treiner}},\ }\href@noop {} {\bibfield  {journal} {\bibinfo  {journal} {Nucl.
  Phys.}\ }\textbf {\bibinfo {volume} {A422}},\ \bibinfo {pages} {103}
  (\bibinfo {year} {1984})}\BibitemShut {NoStop}%
\bibitem [{\citenamefont {{Reinhard}}\ \emph {et~al.}(1999)\citenamefont
  {{Reinhard}}, \citenamefont {{Dean}}, \citenamefont {{Nazarewicz}},
  \citenamefont {{Dobaczewski}}, \citenamefont {{Maruhn}},\ and\ \citenamefont
  {{Strayer}}}]{Reinhard1999}%
  \BibitemOpen
  \bibfield  {author} {\bibinfo {author} {\bibfnamefont {P.~G.}\ \bibnamefont
  {{Reinhard}}}, \bibinfo {author} {\bibfnamefont {D.~J.}\ \bibnamefont
  {{Dean}}}, \bibinfo {author} {\bibfnamefont {W.}~\bibnamefont
  {{Nazarewicz}}}, \bibinfo {author} {\bibfnamefont {J.}~\bibnamefont
  {{Dobaczewski}}}, \bibinfo {author} {\bibfnamefont {J.~A.}\ \bibnamefont
  {{Maruhn}}}, \ and\ \bibinfo {author} {\bibfnamefont {M.~R.}\ \bibnamefont
  {{Strayer}}},\ }\href@noop {} {\bibfield  {journal} {\bibinfo  {journal}
  {Phys. Rev. C}\ }\textbf {\bibinfo {volume} {60}},\ \bibinfo {eid} {014316}
  (\bibinfo {year} {1999})}\BibitemShut {NoStop}%
\bibitem [{\citenamefont {Roca-Maza}\ \emph {et~al.}(2012)\citenamefont
  {Roca-Maza}, \citenamefont {Colo},\ and\ \citenamefont {Sagawa}}]{Roca2012}%
  \BibitemOpen
  \bibfield  {author} {\bibinfo {author} {\bibfnamefont {X.}~\bibnamefont
  {Roca-Maza}}, \bibinfo {author} {\bibfnamefont {G.}~\bibnamefont {Colo}}, \
  and\ \bibinfo {author} {\bibfnamefont {H.}~\bibnamefont {Sagawa}},\
  }\href@noop {} {\bibfield  {journal} {\bibinfo  {journal} {Physical Review
  C}\ }\textbf {\bibinfo {volume} {86}},\ \bibinfo {pages} {031306} (\bibinfo
  {year} {2012})}\BibitemShut {NoStop}%
\bibitem [{\citenamefont {Erler}\ \emph {et~al.}(2013)\citenamefont {Erler},
  \citenamefont {Horowitz}, \citenamefont {Nazarewicz}, \citenamefont
  {Rafalski},\ and\ \citenamefont {Reinhard}}]{Erler2013}%
  \BibitemOpen
  \bibfield  {author} {\bibinfo {author} {\bibfnamefont {J.}~\bibnamefont
  {Erler}}, \bibinfo {author} {\bibfnamefont {C.J.}~\bibnamefont {Horowitz}},
  \bibinfo {author} {\bibfnamefont {W.}~\bibnamefont {Nazarewicz}}, \bibinfo
  {author} {\bibfnamefont {M.}~\bibnamefont {Rafalski}}, \ and\ \bibinfo
  {author} {\bibfnamefont {P.-G.}\ \bibnamefont {Reinhard}},\ }\href@noop {}
  {\bibfield  {journal} {\bibinfo  {journal} {Phys. Rev. C}\ }\textbf {\bibinfo
  {volume} {87}},\ \bibinfo {pages} {044320} (\bibinfo {year}
  {2013})}\BibitemShut {NoStop}%
\end{thebibliography}%
\end{document}